\documentclass[conference]{IEEEtran}
\usepackage[colorlinks, urlcolor=blue, citecolor=blue]{hyperref}

\usepackage{amsmath}
\usepackage{subcaption}
\usepackage{kantlipsum} 
\usepackage{mwe}
\usepackage{adjustbox}
\usepackage{wrapfig}
\usepackage{float}
\usepackage{subcaption}
\usepackage[ruled,linesnumbered]{algorithm2e}

\SetAlFnt{\small}
\SetAlCapFnt{\small}
\SetAlCapNameFnt{\small}
\SetAlCapHSkip{0pt}
\IncMargin{-\parindent}
\newcommand{\eg}{{\it e.g., }}
\newcommand{\etal}{{\it et~al. }}
\newcommand{\ie}{{\it i.e., }}
\newcommand{\name}{SAED}
\usepackage{fixltx2e}
\usepackage{flexisym}
\usepackage{multirow}
\usepackage{mathtools, setspace,lscape,algorithmic,verbatim,color, alltt, amsfonts, wrapfig, boxedminipage,pifont,mathtools }
\usepackage{dblfloatfix}
\usepackage{aliascnt}

\usepackage{xcolor}
\usepackage{balance}
\usepackage{fancyhdr, lastpage}
\usepackage{subcaption}
\begin{document}
\title{\name: Edge-Based Intelligence for Privacy-Preserving Enterprise Search on the Cloud}

\author{\IEEEauthorblockN{Sakib M Zobaed\IEEEauthorrefmark{1},
Mohsen Amini Salehi\IEEEauthorrefmark{1}, and Rajkumar Buyya\IEEEauthorrefmark{2}}

\IEEEauthorblockA{\IEEEauthorrefmark{1}High Performance Cloud Computing (HPCC) lab, \\School of Computing \& Informatics,
University of Louisiana at Lafayette, USA \\
}
\IEEEauthorblockA{\IEEEauthorrefmark{2}Cloud Computing \& Distributed
Systems (CLOUDS) lab,
School of Computing and Information Systems\\
The University of Melbourne, Parkville, VIC 3010,
Australia\\Email: \{sm.zobaed1, amini\}@louisiana.edu, rbuyya@unimelb.edu.au
}
}


\maketitle
\begin{abstract}

Cloud-based enterprise search services (\eg AWS Kendra) have been entrancing big data owners by offering convenient and real-time search solutions to them. However, the problem is that individuals and organizations possessing confidential big data are hesitant to embrace such services due to valid data privacy concerns. In addition, to offer an intelligent search, these services access the user's search history that further jeopardizes his/her privacy. To overcome the privacy problem, the main idea of this research is to separate the intelligence aspect of the search from its pattern matching aspect. According to this idea, the search intelligence is provided by an on-premises edge tier and the shared cloud tier only serves as an exhaustive pattern matching search utility. We propose \emph{Smartness at Edge} (\name~mechanism that offers intelligence in the form of semantic and personalized search at the edge tier while maintaining privacy of the search on the cloud tier. At the edge tier, \name~uses a knowledge-based lexical database to expand the query and cover its semantics. \name~personalizes the search via an RNN model that can learn the user's interest. A word embedding model is used to retrieve documents based on their semantic relevance to the search query.
\name~is generic and can be plugged into existing enterprise search systems and enable them to offer intelligent and privacy-preserving search without enforcing any change on them. 
Evaluation results on two enterprise search systems under real settings and verified by human users demonstrate that \name~can improve the relevancy of the retrieved results by on average $\approx24\%$ for plain-text and $\approx75\%$ for encrypted generic datasets.

\end{abstract}


\maketitle

\section{Introduction}\label{sec:intro}

The expeditious growth of digitalization has been producing a massive volume of data, known as \emph{big data}, in both structured and unstructured formats. It is estimated that $95\%$ of the generated data is in the unstructured format, produced from various sources, such as organizational documents, emails, web pages, and social networks~\cite{exa}. Cloud services have been effective in relieving big data owners from the burden of maintaining these data. Recently, cloud providers began offering \emph{enterprise search} services that enable data owners to semantically search over their datasets in the cloud. For instance, AWS has launched an enterprise search service named AWS Kendra~\cite{kendra} that offers real-time semantic searchability using natural language-based machine learning techniques.  

Although the cloud services have been fascinating for big data owners \cite{samani2020art}, there have been numerous privacy violation incidents~\cite{cloudacci} during recent years that have made individuals and businesses with sensitive data (\eg healthcare documents) hesitant to fully embrace the data management cloud services. In one incident, confidential information of over three billion Yahoo users were exposed~\cite{clustcrypt}. In another incident, information of over $14$ million Verizon customer accounts were exposed from the company's cloud system~\cite{clustcrypt}.

Ideally, data owners desire a privacy-preserving cloud service that offers semantic and personalized searchability in a real-time manner, without overwhelming their resource-constrained (thin) client devices (\eg smartphones). A large body of research has been undertaken on privacy-preserving enterprise search services in the cloud \cite{li,sun,amini14,S3C,S3BD} whose goals are to protect user's sensitive data from internal and external attackers. However, most of these works fall short in retrieving search results that are semantically relevant to the context and  user's interest (\ie personalized search)~\cite{S3BD,S3C}.  
In addition, these works often rely on the client device and impose significant overhead on it to perform a secure query processing or to encrypt/decrypt user documents.

To satisfy all of the aforementioned desires of a particular user, our main idea in this research is to separate the intelligence aspect of the enterprise search from its pattern matching aspect. According to this idea, we propose to leverage on-premises edge computing~\cite{zobaedbig,nur2019priority} to handle the search intelligence and user-side encryption. For that purpose, the edge-based mechanism, called \emph{Smartness At Edge (\name)}, is developed to extract both contextualized and personalized semantics from the search query and the user's search history as well.

Then, \name~feeds the cloud resources with  proactively augmented and encrypted search queries. In this case, the high-end cloud resources are employed only to store encrypted contents and to exhaustively perform pattern matching of the fed query across the entire dataset.

\begin{figure}
	\centering
	\includegraphics[width=\linewidth]{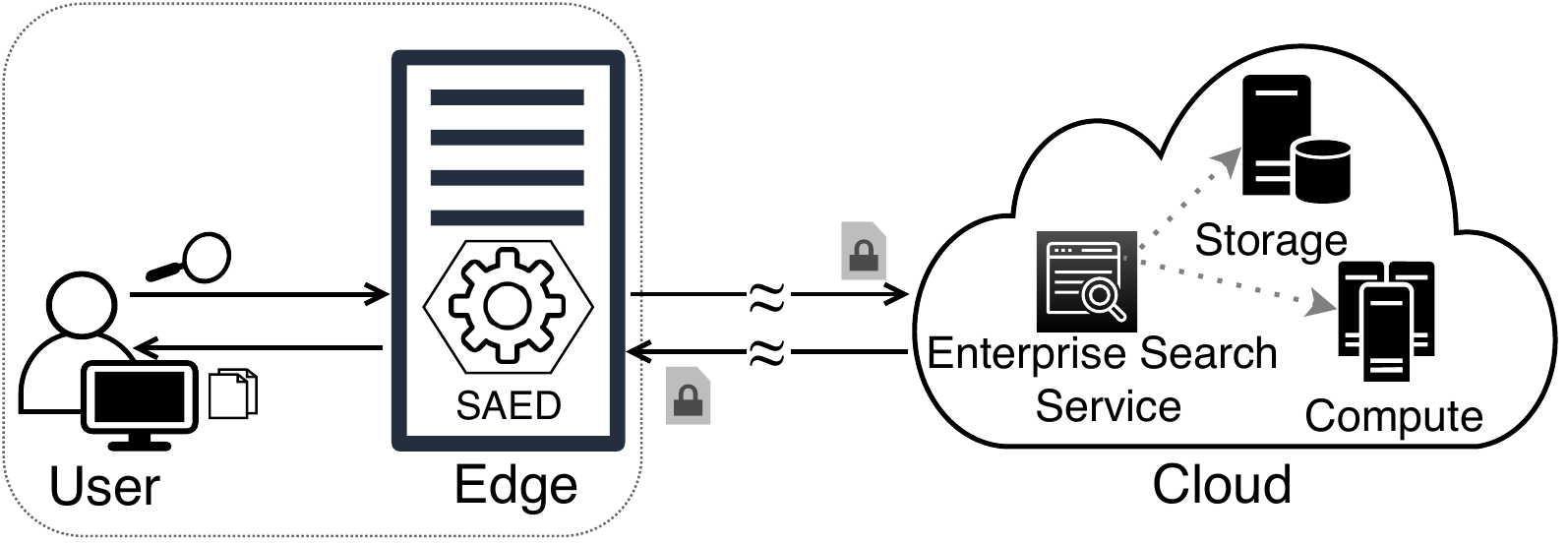}
	\caption{\small{Bird-eye view of \name~mechanism in a three-tier architecture to facilitate smart and privacy-preserving enterprise search service. \name~provides the secure search intelligence on the on-premises edge resources. The high-end storage and compute resources on the cloud tier are utilized by the existing enterprise search systems to exhaustively carry out pattern matching on the entire dataset.}}
	\label{fig: search_ac_high}
\end{figure}

Figure~\ref{fig: search_ac_high} provides a bird-eye view of the \name~mechanism. On one end, it communicates with the client device(s) to handle the security processing of the user contents and to achieve search intelligence before feeding the query set to the cloud tier. On the other end, \name~communicates with the cloud tier where an existing enterprise search service (\eg Kendra) works with the computing and storage services in the cloud to perform exhaustive pattern matching of the encrypted query set on the uploaded dataset. Upon completion of the pattern matching process, the set of resulting documents is retrieved and ordered by \name~with respect to the user's interests. Ultimately, the ordered results are handed over to the user's device.

To identify the actual context of the query and to proactively expand it to a set of contextually-related queries, we leverage WordNet \cite{miller1995wordnet} that is a widely adopted knowledge-based lexical database. However, contextualizing the query cannot help in certain scenarios where the query is short and ambiguous. For instance, considering \texttt{jaguar} as the search query, it can be contextualized to both a car brand or a wild animal. For this type of queries, identifying the user's interest can complement the contextualization and navigate the search towards the semantics intended by the user (\ie achieving personalized search). For that purpose, \name~utilizes a recurrent neural network model to infer the user's interest based on his/her search history.  
Although proactive query expansion (\ie augmenting the user query to a set of queries) is vital to capture the search semantics, not every element of the expanded query set is equally relevant to the original query. As such, \name~assigns a weight to each expanded query that represents its semantic distance to the original query. 

In summary, the contributions of the work are as follows:
\begin{itemize}
\item We develop the open-source \name~mechanism at the edge tier that offers personalized semantic searchability on existing cloud-based enterprise search services while maintaining data privacy. 

\item We propose a method to extract the context of a given search query that often appears in form of a short and incomplete sentence. 

\item We design a method for proactive query expansion to cover the search semantic with respect to its context. 

\item We develop a method based on a recurrent neural network model to personalize the search via assigning a weight to each expanded query.

\item We evaluate the search accuracy and privacy of \name~via plugging it to the existing cloud-based search services.

\end{itemize}

The rest of the paper is organized as follows. In Section \ref{sec-background}, we discuss background study and related prior works.
Later, we provide the architectural details of~\name~mechanism in Section~\ref{sec:propo}. In Section~\ref{sec:kendra}, we discuss the pluggability of~\name~in the context of AWS Kendra. We discuss results and performance analysis in Section \ref{sec:evltn}. Finally, Section \ref{sec:conclsn} concludes the paper.

\section{Background and Prior Literature}
\label{sec-background}

Several research works have been undertaken in semantic and/or privacy-aware search systems. Here, we introduce some notable mentions and position the contributions of \name~against them. 
\begin{figure*} [!htb]

	\centering
	\includegraphics[width=.7\linewidth]{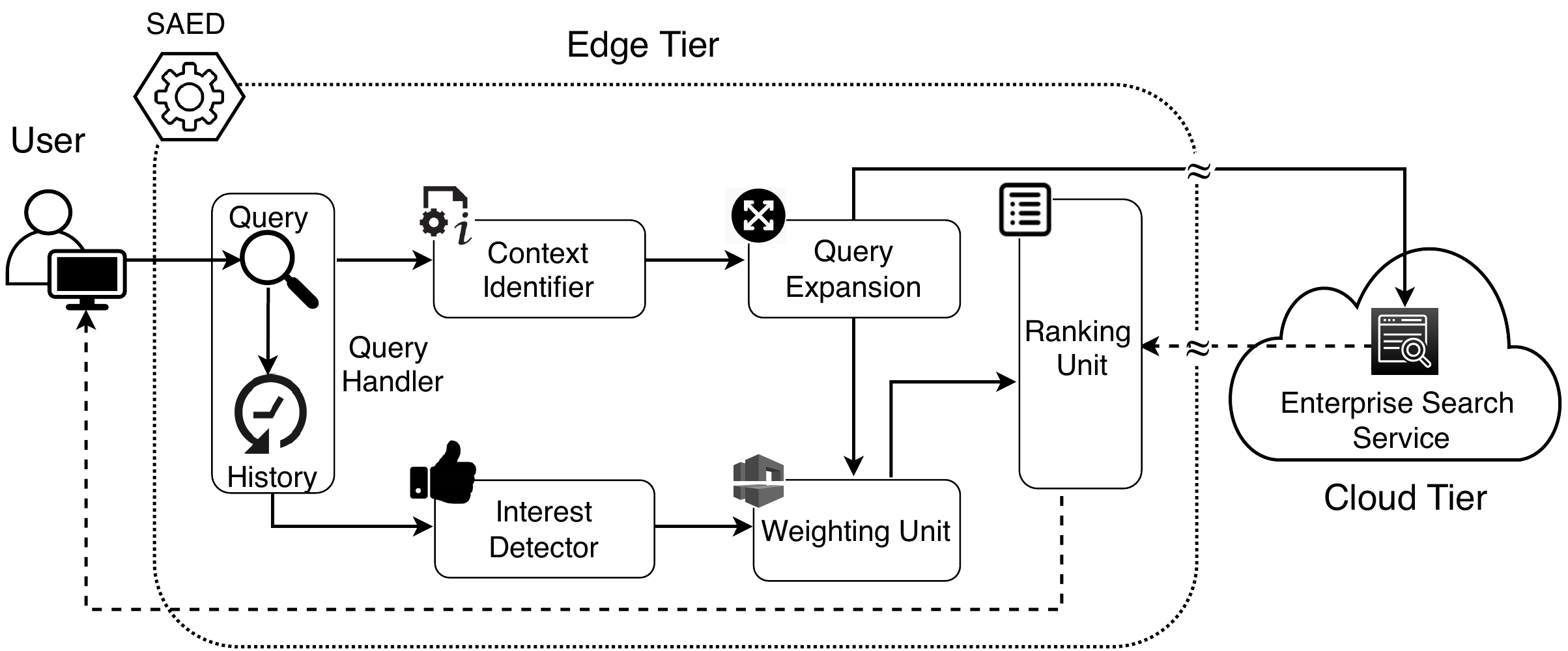}

	\caption{\small{Architectural overview of the \name~system within edge tier and as part of the three-tier enterprise search service. \name~provides semantic search via identifying the query context and combining that with the user's interests. Then, Query Expansion and Weighting unit of \name, respectively, incorporate the semantic and assure the relevancy of the results. Solid and dashed lines indicate the interactions from user to the cloud tier and from the cloud tier to the user respectively. }}
	\label{fig: search_ac}
	
\end{figure*}
\subsection{Cloud-Based Enterprise Search Services}

Cloud-based enterprise search services, such as AWS Kendra, offer semantic searchability, given that they are provided with the plain-text data. That means the semantic ability comes with the cost of compromising the users' data privacy \cite{S3BD,sahan,clustcrypt}. This is, in fact, the trapdoor that particularly internal attackers can misuse to breach the confidentiality or even the integrity of the users' data. \textit{It is this type of attack model that we try to make the cloud-based enterprise search services resistant against.} We note that, for encrypted datasets, the current enterprise search services cannot offer anything beyond na\"ive string matching. 

Even for plain-text datasets, our investigations revealed that Kendra covers only ontological semantics in the search and it falls short in providing context-aware and personalized semantics. 
For instance, we tested Kendra to verify the ability of capturing context-aware semantics by feeding \texttt{soccer} as a query and in the result set, there were documents about \texttt{rugby}. In another test, \texttt{river bank} query returned documents about \texttt{commercial bank} that indicates the lack of context-awareness in the search.  

Alternatively, \name~can offer context-aware and personalized search while maintaining data privacy. It can be plugged into any enterprise search service without enforcing any change on them and enrich their semantic search quality by incorporating context-awareness and personalization.

\subsection{Semantic Representation of Query Keywords}
Query expansion is a process to seek keywords that are semantically related to a given query and fill the lexical gap between the user queries and the searchable documents. One of the widely-used methods of query expansion is Pseudo-Relevance Feedback (PRF)~\cite{liang2012exploiting,wang2019query} that extends an unsuccessful query with various related keywords and then re-ranks the search results to increase the likelihood of retrieving relevant documents. Although the PRF-based approach generally improves the retrieval effectiveness,
it is sensitive to the quality of the original search results.



Latent semantic analysis~\cite{deerwester1990indexing}, latent dirichlet analysis~\cite{albishre2017effective}, and neural-based linguistic models~\cite{wang2019query,diaz2016query} are some of the query expansion methods that can obtain the semantic representation of a given query.    
In these methods, vectors are commonly referred to as \emph{word embeddings} that represent words into a low-dimensional semantic space, where the vicinity of words demonstrates the syntactic or semantic similarity between them~\cite{mikolov2013efficient}. 
However, pre-trained word embedding models, such as Word2vec \cite{mikolov2013efficient}, always generate the same vector representation for an input word, regardless of the context in which the word has appeared in. Hence, if any ambiguous keyword(s) present in a query, the underlying topic of the query could not be detected. 

WordNet \cite{miller1995wordnet} is one of the widely-used and lexically-rich resources in English that is utilized to infer the sense of ambiguous words in a given corpus. 
In WordNet, words containing similar meanings are grouped into synonym sets, whereby each set has a semantic and conceptual relationship with the other sets. Song \etal~\cite{song2007integration} and Nakade~\etal~\cite{nakade2018preliminary} evaluate the 
effectiveness of utilizing WordNet for query expansion in National Institute of Standards and Technology (NIST) and Twitter datasets. They identify important key-phrases of the query and use WordNet to obtain the relevant synonym sets. Later, they utilize the synonym sets to construct the expanded query.
Nevertheless, in most of the prior research on query expansion using WordNet (\eg~\cite{leung2013collective}), the elements of the expanded query set are considered uniformly that undermines the relevancy and ranking of the result set.

\subsection{Privacy-Preserving Search Systems} 

In addition to plain-text data, searching is performed on privacy-preserving data ensuring negligible chances of data leakage. Therefore, various searchable encryption-based solutions are adopted to facilitate search over such data.




Few works at the time of writing have combined the ideas of semantic searching and searchable encryption. Works that attempt to provide a semantic search often only consider word similarity instead of true semantics.  
Li~\etal~\cite{li} propose a system which could handle minor user typos through a fuzzy keyword search. Moataz~\etal~\cite{moataz} use various stemming approaches on terms in the index and query to provide more general matching. Sun~\etal~\cite{sun} present a system that used an indexing approach over encrypted file metadata and data mining techniques to capture the semantics of queries. This approach, however, builds a semantic network only using the documents that are given to the set and only considers words that are likely to co-occur as semantically related, leaving out many possible synonyms or categorically related terms.  
Woodworth~\etal propose S3BD~\cite{S3BD}, a secure semantic search system that could search semantically over encrypted confidential big data. 
They expand their search query by incorporating semantic data extracted blindly from an ontological network.
They do not consider context-aware query expansion that created confusion for the search system while processing ambiguous or multi-context keywords in a query. To perform query processing in client devices, they end up requiring additional computational overhead in the client tier. 



 \section{\name: Smart Edge-Based Enterprise Search System}
\label{sec:propo}
\subsection{Architectural Overview}
In this part, we provide a bird-eye view of the \name~system, that enables intelligent and secure enterprise search in the cloud. The system is structured around three tiers, shown in Figure~\ref{fig: search_ac_high}, and explained as follows:

\begin{itemize}
    \item \emph{Client tier} (\eg smartphone, tablet) contains a lightweight application that provides a user interface for uploading documents and to search over
them in the cloud. Datasets are either uploaded by the user or by the organization that owns the data. 

\item \emph{Edge tier} extracts representative keywords of the documents being uploaded to the cloud tier and builds an index on the cloud tier. Upon receiving a search query from the client tier, the \name~system on the edge tier offers intelligence by considering the query semantics and the user's interest. The edge tier is located in the client's premises, hence, deemed as an honest and secure system. To offer a secure enterprise search service, the edge tier encrypts both the uploaded data and the search query. In addition, it decrypts the result set before delivering it back to the client tier. 

\item \emph{Cloud tier} contains numerous high-end servers that are utilized for storing (encrypted) data and performing the large-scale computation required to exhaustively search against the index \cite{S3C,S3BD}. The index can be clustered based on the underlying topics of its keywords (please refer to our prior works~\cite{S3BD,clustcrypt} for further details).                   
    
\end{itemize}



In Figure~\ref{fig: search_ac}, we depict the components of \name~and show the interactions between them. At first, a user-provided search query is received by the \emph{Query Handler} that keeps track of the user's search history and initializes the \emph{Context Identifier} unit whose job is to extract the context and disambiguate the query phrase. Then, according to the extracted context, the query is proactively expanded by the \emph{Query Expansion} unit and a \emph{query set} is constructed. To achieve the personalized search, the \emph{Interest Detector} unit of \name~leverages the user's search history to recognize his/her interest and weight each element of the query set (\ie expanded queries) based on its relatedness to the user interest. Once the pattern matching phase is accomplished on the cloud tier, the resulted documents are returned to \name~on the edge tier. Next, the \emph{Ranking Unit} utilizes the assigned weights to order the retrieved documents based on their relevance to the user's interest and generates a retrieved document list, denoted as $D_\theta$, that is sent to the user's device. 
In the next parts, we elaborate on each unit of the \name~system.

\subsection{Query Context Identification}
Identifying the context of a given search phrase is vital to navigate the search to the semantics intended by the user. Considering the example of \texttt{cloud computing} as the search query, without a proper context identification the returned document set can potentially include documents about \texttt{sky} and \texttt{climate}, whereas, an efficient context identifier can recognize the right semantic and navigate the search to the topics around \texttt{distributed, edge, fog,} and \texttt{cloud computing}. In fact, identifying the context helps the Query Expansion unit to form a query set diversified around relevant keywords that semantically represent the search query and subsequently improve the relevancy of the results.

Prior context identification works (\eg \cite{silva2020improving,kuzi2016query, diaz2016query}) have the following shortcomings: \emph{first,} they often assume each keyword has the same importance in the query and recognize the query context via averaging the embeddings of its keywords. However, not all keywords in a query necessarily help in identifying the context. For example, the keyword \texttt{various} in \texttt{various cloud providers} does not bring any significance to the context and can be eliminated. \emph{Second,} the embedding methods used by the existing works always provide the same representation for a given keyword, irrespective of the underlying context. This is particularly problematic for ambiguous keywords whose meaning vary based on the query context. For instance, the embedding of \texttt{cloud} in the aforementioned example should be different when it is used along with the \texttt{computing} as opposed to when it is used along with the \texttt{weather} in a given query. \emph{Third,} existing methods only consider the embeddings of the common keywords, while discarding most of the name-entities (\eg names and locations) that do not exist in the vocabulary of Word2Vec \cite{miller1995wordnet,fellbaum2017wordnet}. For instance, consider \texttt{best selling books of J.K. Rowling} as the query; \texttt{Book} and \texttt{Sell} are identified as the query context and \texttt{J.K. Rowling} is discarded. However, our analysis suggests that the context of a short query phrase often has contextual association with the discarded name-entities. 

To overcome the shortcomings and identify the actual context of a given query, we propose to take a holistic approach and extract the \emph{semantic across query keywords, proportionate to the importance of each keyword}. The main output of the Context Identification unit is a set of keywords, denoted as \emph{C}, that collectively represent the context of the query.

Specifically, to eliminate unimportant keywords that do not contribute to the semantic of query $Q$, the Context Identification unit utilizes \emph{Yake}~\cite{yake}, which is a unsupervised keyword extractor that discards unimportant keywords of the query. The remaining keywords (\ie the trimmed query, denoted as the $Q^\prime$ set) are considered for context identification. To learn the true semantic of $Q^\prime$, the unit leverages the Lesk algorithm \cite{fellbaum2017wordnet} of WordNet to disambiguate each keyword $q\in Q^\prime$. Lesk algorithm works based on the fact that keywords in a given sentence (query) tend to imply a certain topic. For keyword $q$, Lesk can determine its true semantics via comparing the dictionary definitions of $q$ against other keywords in $Q^\prime$ (\ie $Q^\prime - \{q\}$). Let $c_q$ be the set of keywords representing the context of $q$. Then, the context of $Q$ is determined as $C=\cup_{\forall q\in Q^\prime} c_q$. Lastly, the Context Identifier recognizes name-entities from $Q$ using WordNet and considers them as part of the context, but in a separate set, denoted as $N$. The reason for considering a separate set is that we apply a different treatment on $N$ and $C$ in the other units of \name.

\begin{algorithm}
	\SetAlgoLined\DontPrintSemicolon
	\SetKwInOut{Input}{Input}
	\SetKwInOut{Output}{Output}
	\SetKwFunction{algo}{algo}
	\SetKwFunction{proc}{Procedure}{}{}
	\Input{query $Q$}
	\Output{$C$: set of keywords representing context of $Q$, \\$N$: set of name-entity in $Q$ }

	\SetKwBlock{Function}{Function \texttt{contextIdentification($Q$):}}{end}
	
	\Function{
	$Q^\prime \gets$ extract keywords from $Q$ using Yake alg.\;
	
	\ForEach {$q \in Q$}{ 
	  
	  \If {$q \in$ {Name-entity}}
	  {
	    $N\gets N \cup \{q\}$ \;
	  }
	  \Else
	  {
	   
	    \If{$q \in Q^\prime $} { 
	    $E_q\gets$ define $q$ based on $Q^\prime -q$ using Lesk alg.\;
	    $c \gets$ extract set of keywords of $E_q$ using Yake alg. \;
        $C \gets C \cup c$\;
	   } 
	  }
	  
	  }
	   
	  return  {$C, N$} \;
	}
	\caption{Pseudo-code to detect the context of a given query in the Context Identification unit of \name.}
	\label{alg:con_iden}
	\end{algorithm}
	
Algorithm~\ref{alg:con_iden} provides a pseudo-code for identifying the context of incoming query $Q$. The outputs of the pseudo-code are two sets, namely $C$ and $N$, that collectively represent the context of $Q$. In Step 2 of the pseudo-code, Yake algorithm is used to filter $Q$ by extracting its important keywords and generate the $Q^\prime$ set.
Name-entities of $Q$ are identified by checking against WordNet and form the set $N$ (Steps 4--6).
Next, in Steps 8--12, for each keyword $q\in Q^\prime$, the Lesk algorithm is employed to disambiguate $q$ and find its true definition with respect to the rest of keywords in $Q^\prime$. Important keywords of the definitions form the context set ($C$) for $Q$.  


\subsection{Query Expansion Unit}
\label{query_expansion}
The \emph{Query Expansion} unit is in charge of proactively expanding the query keywords based on their relevant synonyms that are in line with their identified context. Neglecting the query context and blindly considering all the synonyms, as achieved in \cite{silva2020improving, kuzi2016query, diaz2016query, S3BD},  leads to finding irrelevant documents. Accordingly, the unit leverages the context of $Q$ (\ie $C$ and $N$) to only find the set of synonyms, denoted as $P$, that are semantically close to the query context. 

Word2Vec \cite{mikolov2013efficient} is a shallow neural network model that can be trained to generate vector representation of keywords, such that the cosine similarity of two given keywords indicates the semantic similarity between them. Accordingly, to proactively expand each keyword $q\in Q$, the Query Expansion unit instruments Word2Vec, pre-trained with Google News dataset \cite{khatua2019tale}, to form the set of nominated synonyms, denoted as $s_q$. 
Let $s_q^i$ be a synonym of $q$ (\ie $s_q^i\in s_q$). Then, the similarity of $s_q^i$ and the query context, denoted as $sim(s_q^i,C)$, is defined based on the sum of similarities with each element of $C$, as shown in Equation~\ref{eq:queryexp}.

\begin{equation}
\label{eq:queryexp}
    sim(s_q^i , C)=\sum_{\forall C_{j}\in C}sim(s_q^i,C_j)
\end{equation}

Then, $s_q^i$ is chosen as an element of $P$, only if it is semantically close enough to the query context. To determine the sufficient closeness, we consider $sim(s_q^i,C)$ to be greater than the mean of the pair-wise similarity across all members of $s_q$ (\ie $sim(s_q^i,C) > \mu_{\forall q \forall j}(sim(s_q^j,C))$). We note that because the elements of $C$ and $N$ represent the context of $Q$, they as well are added to $P$.

Algorithm~\ref{alg:queryExpansion} provides a high level pseudo-code for generating the expanded query set $P$. In Steps 2--7 of the pseudo-code, the synonym set for each $q$ is generated. Next, the similarity between each word $s_q^i$ and $C$ is calculated. The similarity values are used to calculate the mean similarity of all nominated queries in Step 8. In Steps 9--15, expanded query set $P$ is formed by including nominated synonyms whose semantic closeness is greater than $\mu$. Lastly, in Step 16, set $P$ is expanded by including context set and name-entities.

\begin{algorithm}
	\SetAlgoLined\DontPrintSemicolon
	\SetKwInOut{Input}{Input}
	\SetKwInOut{Output}{Output}
	\SetKwFunction{algo}{algo}
	\SetKwFunction{proc}{Procedure}{}{}
	\SetKwFunction{main}{\textbf{ChooseCenter}}
	\Input{$Q$, $C, N$}
	\Output{$P$: the expanded query set  }
	
	\SetKwBlock{Function}{Function \texttt{ QueryExpansion($Q$, $C$, $N$)}}{end}
	
	\Function{
    	\ForEach{$q\in Q$} {
            $s_q \gets$ use WordNet to obtain synonym set of $q$ \;  
            \ForEach {$s_q^i \in s_q$} {
    	        $sim(s_q^i,C) \gets \displaystyle\sum_{\forall C_{j}\in C}sim(s_q^i,C_j)$ \;
    	}
    	}
         $\mu\gets$ calculate mean $sim(s_q^j,C)$ across all $q\in Q, \forall s_q^j\in s_q$ \;
        \ForEach{$q\in Q$}{
            \ForEach {$s_q^i \in s_q$} {
             \If{$sim(s_q^i,C) > \mu$}  
                {Add $s_q^i$ to set $P$}
            }  
        }
        $P \gets P \cup C \cup N$ \;
	  return  $P$ \;
	}
	\caption{Pseudo-code to expand query based on the context in the Query Expansion unit of \name}
	\label{alg:queryExpansion}
	
\end{algorithm}

\subsection{User Interest Detection}  
Detecting the user's search interest is essential to deliver personalized search. In \name, interest detection is achieved by analyzing two factors: (A) the user's search history; and (B) the user's reaction to the retrieved results of prior search queries. This can be detected based on the results chosen by the user or the time spent for browsing them.  

Let $\Delta^\prime$ represent the whole resulted documents that are sent to the user and $\tau$ represent the documents where the user is interested in. We have $\tau \subseteq \Delta_\prime$. Accordingly, the user's interest can be derived from the topics of $\tau$.  
The Interest Detector unit uses an existing document classification model \cite{kastrati2019impact}, operating based on Na\"ive Biased (NB) method, to determine the topics of $\tau$, denoted as $t_\tau$. We also perform majority voting on $t_\tau$ to find the user's main interest. The process is repeated to store \textit{n}-prior search interests data of the user. The data is characterized as sequential as it is harvested from each successful search.  
By analyzing the user's prior search interests, the edge tier trains a recurrent neural network-based prediction model~\cite{pang2020innovative} that can predict the user's search interest. In case of \name, as the data does not contain long dependency and to keep the model simple and to maintain real-timeliness, instead of a stacked (\ie deeper) model, we feed the harvested user-specific historical search data to train a many-to-one vanilla RNN model~\cite{rnn2019}.

\subsection{Weighting Unit}
Once \name~learns the user interest, the next step to accomplish a context-aware and personalized enterprise search is to determine the closeness of contextually-expanded queries (\ie elements of $P$) to the user's interest. In fact, not all expanded queries have the same significance in the interpretation of the query. Accordingly, the objective of the \emph{Weighting unit} is defined as quantifying the closeness of each expanded query to the user's interest. 
Later, upon completion of the search operation on the cloud tier, the weights are used by the \emph{Ranking unit} of \name~to prune and sort the result set. 

Prior weighting schemes (\eg~\cite{S3C,S3BD,diaz2016query,wang2019query,kuzi2016query}) often use the word frequency-based approach (\eg TF-IDF~\cite{S3BD}) and discard the user interests. Alternatively, the weighting procedure of \name~quantifies the importance of each expanded query $p\in P$ based on two factors: (A) The \emph{type} of $p$, which means if it directly belongs to the context ($C$ and $N$ sets) or is derived from them; and (B) The \emph{semantic similarity} of $p$ to the user interest. 

In particular, those elements of $P$ that directly represent the query context or name-entities (\ie $\forall p | p\in P\cap (C \cup N)$) explicitly indicate the user's search intention, hence, weighting them should be carried out irrespective of the user interest. A deeper analysis indicates that name-entities that potentially exist in a query represent the search intention, thus, biasing the search results to them can lead to a higher user satisfaction. As such, the highest weight is assigned to $\forall p | p \in (P \cap N$). The highest weight is determined by the domain expert, however, in the experiments we consider it as $\eta_{max}=1$. 
We define the \textit{contribution} of $q\in Q$ as the ratio of the number of keywords added to $C$ because of $q$ (denoted $C_q$) to the cardinality of $C$. Let $\eta_p$ denote the weight of $p\in P$. 
Then, for those elements of $P$ that are in the query context (\ie $\forall p \in (P \cap C)$), $\eta_p$ is calculated based on the contribution of the query keyword $q$ corresponding to $p$. Equation~\ref{eq:weight2} formally represents how $\eta_p$ is calculated.
\begin{equation}\label{eq:weight2}
    \eta_p=\frac{\eta_{max}\cdotp |C_q|}{|C|}
\end{equation}

The weight assignment for those $p$ that are derived from elements of $C$, as explained in Section~\ref{query_expansion}, (\ie $\forall p| p\in P-(C \cup N)$) is carried out via considering semantic similarity of $p$ with the user interest $\theta$. That is, $\eta_p=sim(p,\theta)$.

\subsection{Ranking Unit}
Once the expanded query set $P$ is formed, the cloud tier performs string matching for each $p\in P$ across the index structure. We note that, if the user chooses to perform a secure search, the elements of $P$ are encrypted before delivered to the cloud tier. In addition, in our prior works \cite{clustcrypt}, we proposed methods for the cloud tier to cluster the index structure and perform the pattern matching only on the clusters that are relevant to the query. 

The cloud tier returns the resulted document set, denoted as $\Delta$, to the edge tier where the Ranking unit of \name~ranks them based on the relevance and the user's interest and generates a document list, called $\Delta^\prime$ to show to the user. 
For a document $\delta_i \in \Delta$, the ranking score, denoted as $\gamma_i$, is calculated by aggregating the importance values of each $p \in P$ within $\delta_i$ and with respect to its weight ($\eta_p$). The importance of $p$ in $\delta_i$ is conventionally measured based on the \emph{TF-IDF} score \cite{viegas2019cluwords}. Accordingly,   
$\gamma_i$ is formally calculated based on Equation~\ref{eq:ranking engine}. 
\begin{equation}
\label{eq:ranking engine}
\gamma_i= \displaystyle\sum_{\forall p \in P}\Bigg(\eta_p \cdot TFIDF(p, \delta_i )\Bigg)
\end{equation}

The TF-IDF score of $p$ in $\delta_i$ is defined based on the frequency of $p$ in $\delta_i$ versus the inverse document frequency of $p$ across all documents in $\Delta$. Details of calculating the tf-idf score can be found in \cite{viegas2019cluwords}.



Once the Ranking unit calculates the ranking score for all $\delta_i \in \Delta$, then the documents are sorted in the descending order based on their ranks and thus, the document list $\Delta^\prime$ are formed with each$\delta_i$ and displayed to the user. 



\section{\name~As a Pluggable Module to Enterprise Search Solutions}
\label{sec:kendra}

The advantage of \name~is to be independent from the enterprise search service deployed on the cloud tier. That is, using \name~neither interferes with nor implies any change on the cloud-based enterprise search service. \name~can be plugged into any enterprise search solution. It provides the search smartness on the on-premises edge tier and leaves the cloud tier only for large-scale  pattern matching. The whole \name~solution reforms the enterprise search to be semantic, personalized, and confidential services.

In this work, we set \name~to work both with AWS Kendra and S3BD. In the case of using AWS Kendra, the Query Expansion unit sends the expanded query set $P$ to Kendra to search each keyword $p$ against the dataset on the Amazon cloud. The resulted documents are received by \name~and ranked before being delivered to the client tier. In the implementation, we only show top 10 documents from the resulted list to the user. Similarly, we plugged \name~to S3BD to perform confidential semantic search on the cloud. Because S3BD maintains an encrypted index structure that has to be traversed against each search query, the elements of $P$ had to be encrypted before handing them over to the cloud tier. We also verified \name~when it is used along with AWS Kendra where the dataset was encrypted. We noticed that \name~can achieve smart search even when Kendra is set to work with encrypted dataset. The performance measurement and analysis of using \name~along with AWS Kendra and S3BD are elaborated in the next Section. 

\section{Performance Evaluation}\label{sec:evltn}
\subsection{Experimental Setup} 

We have developed a fully working version of \name~and made it available publicly in our Github\footnote{\url{https://github.com/hpcclab/SAED-Security-At-Edge}} page. To conduct a comprehensive performance evaluation of \name~on the enterprise search solutions, we developed it to work with both S3BD~\cite{S3BD} and AWS Kendra~\cite{kendra}. S3BD already has the query expansion and weighting mechanisms, but we deactivated them and set it to use the expanded queries generated by \name. In the experiments, the combination of \name~and S3BD is shown as \name+S3BD. Likewise, the combination of \name~and AWS Kendra is shown as \name+Kendra.

We evaluated \name~using two different datasets, namely \texttt{Request For Comments (RFC)} and \texttt{BBC} that have distinct properties and volume. The reason we chose the \texttt{RFC} dataset is that it is domain-specific and includes $4,951$ documents about the Internet and wireless communication network. Alternatively, the \texttt{BBC} dataset is more diverse. It includes $2,224$ news documents in five distinct categories, including politics, entertainment, business, sports, and technology. 

To conduct a comprehensive evaluation, we used both systematic metrics and human-based feedback as elaborated in Section~\ref{subsec:eval met}. We deployed and experimented \name~on a Virtual Machine (VM) within our local edge computing system. The VM had two 10-core 2.8 GHz E5 Xeon processors with 64 GB memory and Ubuntu 18.4 operating system.

\subsection{Benchmark Queries}
The datasets that we use to carry out the experiments are not featured with any benchmark. Therefore, we required to develop benchmark queries for the datasets before evaluating the performance of \name. We developed $10$ benchmark queries, shown in Table~\ref{tab:benchmark}, for each one of the two datasets. The benchmark queries are proactively designed to explore the breadth and depth of the datasets in question. In addition, some of the queries intentionally contain ambiguous keywords to enable us examining the context detection capability of \name.
For the sake of brevity, we provide one acronym for each benchmark query (see Table~\ref{tab:benchmark}).
For each benchmark query, we collected at most the top-20 retrieved documents. Then, the quality of the retrieved documents were measured via both automated script and human-based users. 

\begin{table}[]

\caption{\small{Benchmark search queries developed for the RFC and BBC datasets.}}
\resizebox{\linewidth}{!}{%
\begin{tabular}{|l|l|}
\hline
\textbf{BBC Dataset}                  & \textbf{RFC Dataset}               \\ \hline
European Commission (\texttt{EC})          & Network Information (\texttt{NI})        \\ \hline
Parliament Archives (\texttt{PA})          & Host Network Configuration (\texttt{HNC}) \\ \hline
Top Camera Phones 2020 (\texttt{TCP})       & Data Transfer (\texttt{DT})              \\ \hline
Credit Card Fraud   (\texttt{CCF})          & Service Extension(\texttt{SE})          \\ \hline
Animal Welfare Bill (\texttt{AWB})          & Transport Layer (\texttt{TL})           \\ \hline
Piracy and Copyright Issues (\texttt{PCI})   & Message Authentication  (\texttt{MA})   \\ \hline
Car and Property Market (\texttt{CPM})      & Network Access (\texttt{NA})             \\ \hline
Rugby Football League  (\texttt{RFL})       & Internet Engineering  (\texttt{IE})      \\ \hline
Opera in Vienna  (\texttt{OV})             & Fibre Channel (\texttt{FC})             \\ \hline
Windows Operating System (\texttt{WOS})     & Streaming Media Service (\texttt{SMS})   \\ \hline
\end{tabular}
}
\label{tab:benchmark}
\end{table}

\subsection{Evaluation Metrics}
\label{subsec:eval met}
We have to measure the search relevancy metric to understand how related the resulted documents are with respect to the user's query and how they meet the his/her interests. For the measurement, we use TREC-Style Average Precision (TSAP) score, described by Mariappan \etal~\cite{mariappan}. TSAP provides a qualitative score in a relatively fast manner and without the knowledge of the entire dataset~\cite{S3BD}.
It works based on the precision-recall concept that is commonly used for judging text retrieval systems. The TSAP score is calculated
based on $\sum_{i=0}^N r_i/N$, where $r_i$ denotes score for $i^{th}$ retrieved document  and $N$ denotes the cutoff number (total number of retrieved documents). Since we consider $N=10$, we call the scoring metric as \emph{TSAP@10}.

To determine $r_i$ for retrieved document $\delta^\prime_i\in \Delta^\prime$, we conducted a human-based evaluation.
We engaged five volunteer students to judge the relevancy of each retrieved document. For every search query, the volunteers labeled each retrieved document as \texttt{highly relevant, partially relevant,} or \texttt{irrelevant}. After performing majority voting based on the provided responses for document $i$, the value of $r_i$ is determined as follows:
\begin{itemize}
\item $r_i = 1/i$ if a document is \texttt{highly relevant}
\item $r_i = 1/2i$ if a document is \texttt{partially relevant}
\item $r_i = 0$ if a document is  \texttt{irrelevant}
\end{itemize}
 
We report TSAP@10 score to show the relevancy of results for each benchmark query. In addition, mean TSAP score is reported to show the overall relevancy across each dataset. As we set the top 10 documents to be retrieved for each search, the highest possible for \emph{TSAP@10} score can be 0.292~\cite{mariappan}. 

In addition to the TSAP score, we measure \emph{Mean F-1} score too to compare the search quality offered by the \name-plugged enterprise search solutions against the original enterprise search solutions (\ie without \name~in place). The F-1 score maintains a balance between the precision and recall metrics, which is useful for unstructured datasets with non-uniform topic distribution.  

\subsection{Evaluating Search Relevancy } 
The purpose of this experiment is to evaluate the search relevancy of enterprise search systems that have \name~plugged into them and compare them against the original (unmodified) systems. To evaluate the personalized search, we set (assumed) \texttt{technology} as the user's interest for both datasets. We note that, in this part, the enterprise search solutions (S3BD and AWS Kendra) are set to work in the plain-text datasets.

\begin{figure*} [h]
\begin{subfigure} {.45\textwidth}
\includegraphics[width=.95\linewidth]{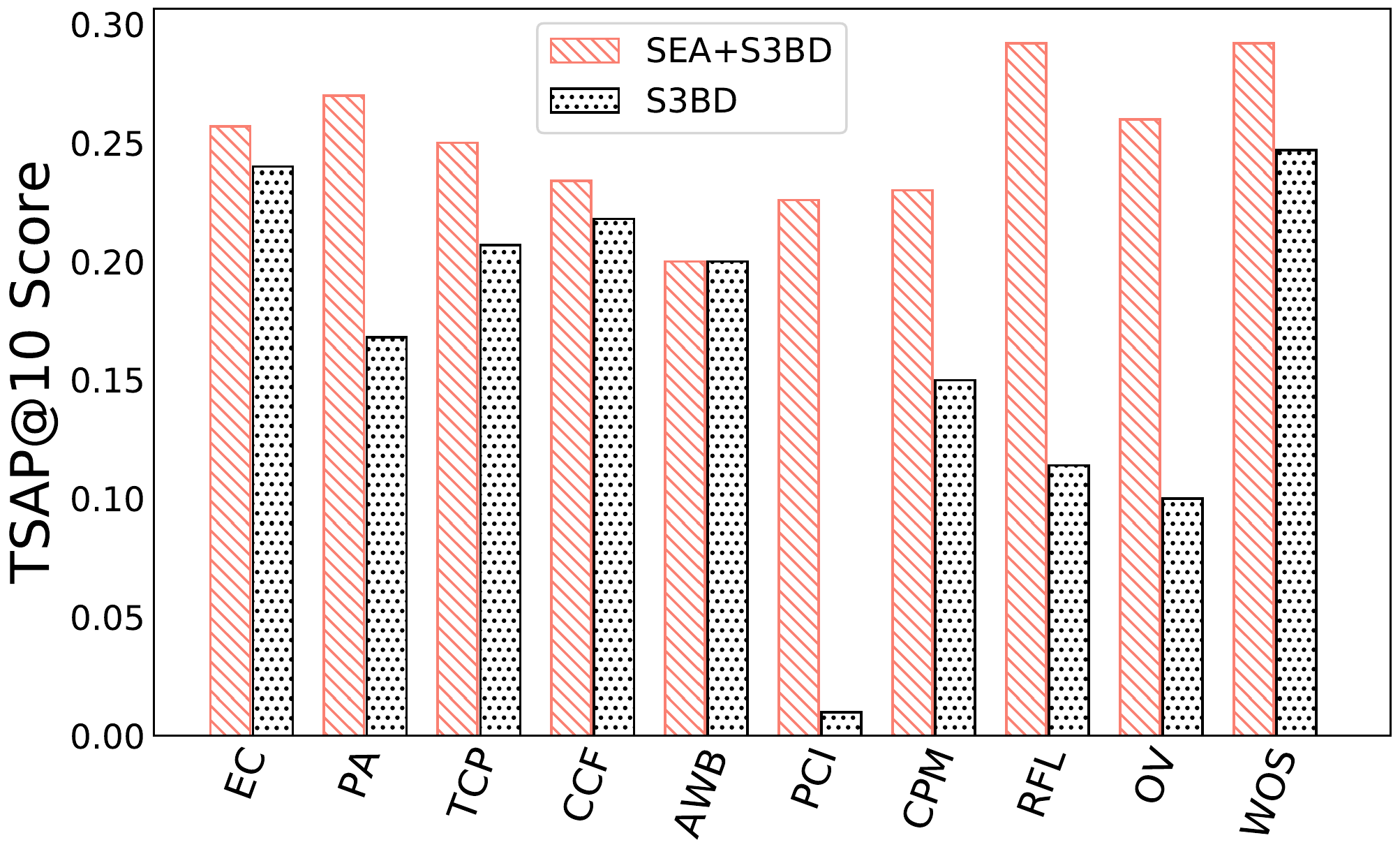}
\caption{BBC dataset}
\label{fig: s3bd bbc}
\end{subfigure} \hfill
\begin{subfigure} {.45\textwidth}
\includegraphics[width=.95\linewidth]{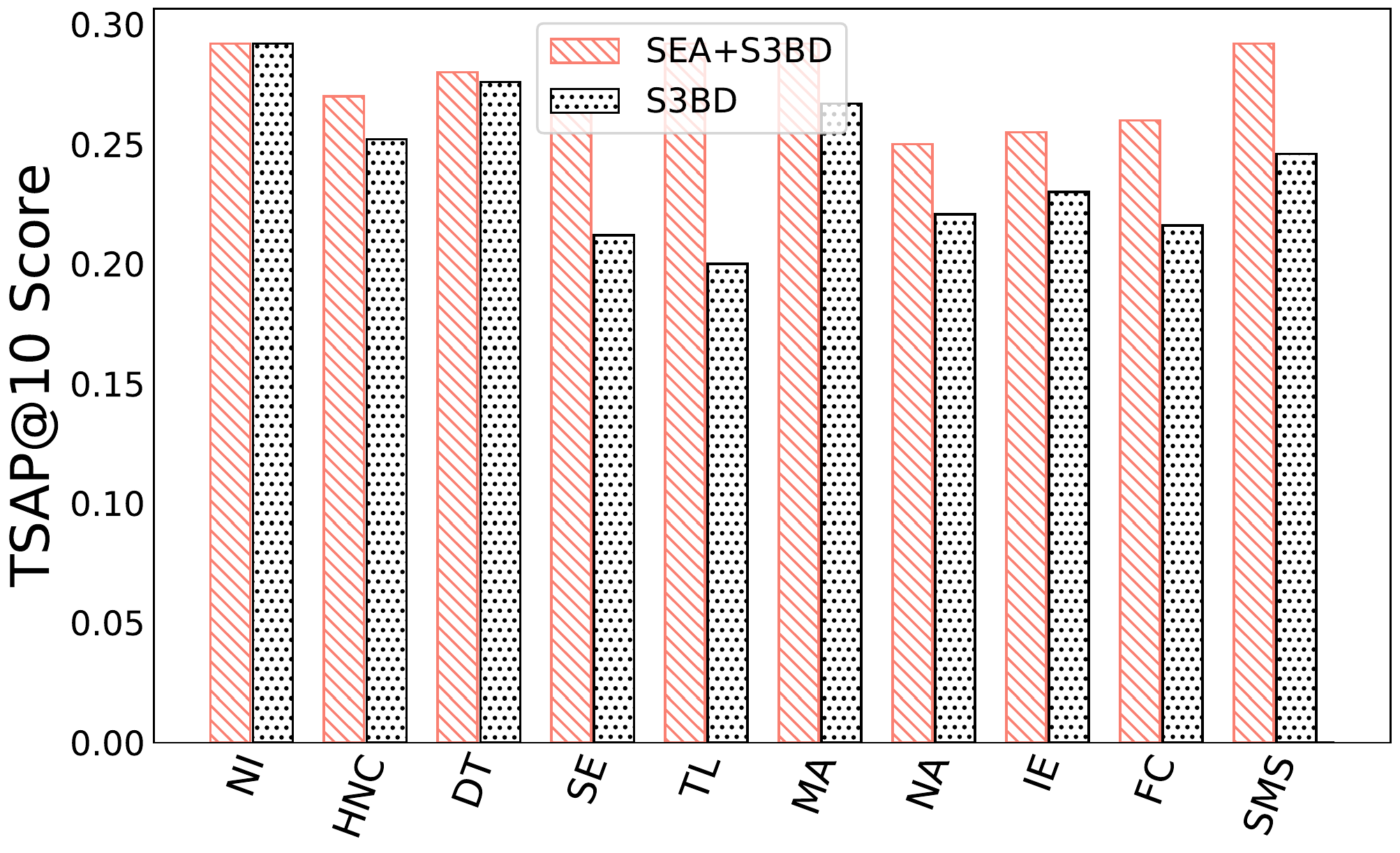}
\caption{RFC dataset}
\label{fig: s3bd rfc}
\end{subfigure} 
\caption{Comparing TSAP@10 scores of \name+S3BD and S3BD systems. Horizontal axes show the benchmark queries.}
\end{figure*}

\paragraph{\textbf{S3BD vs \name+S3BD}}
Figure~\ref{fig: s3bd bbc} shows the TSAP@10 score for the RFC and BBC datasets for the original S3BD and \name+S3BD. The horizontal axes in both subfigures show the benchmark queries and the vertical axes show the search relevancy based on the TSAP@10 score. 

\begin{figure*} [h]
\begin{subfigure}  {.45\textwidth}
\includegraphics[width=.95\linewidth]{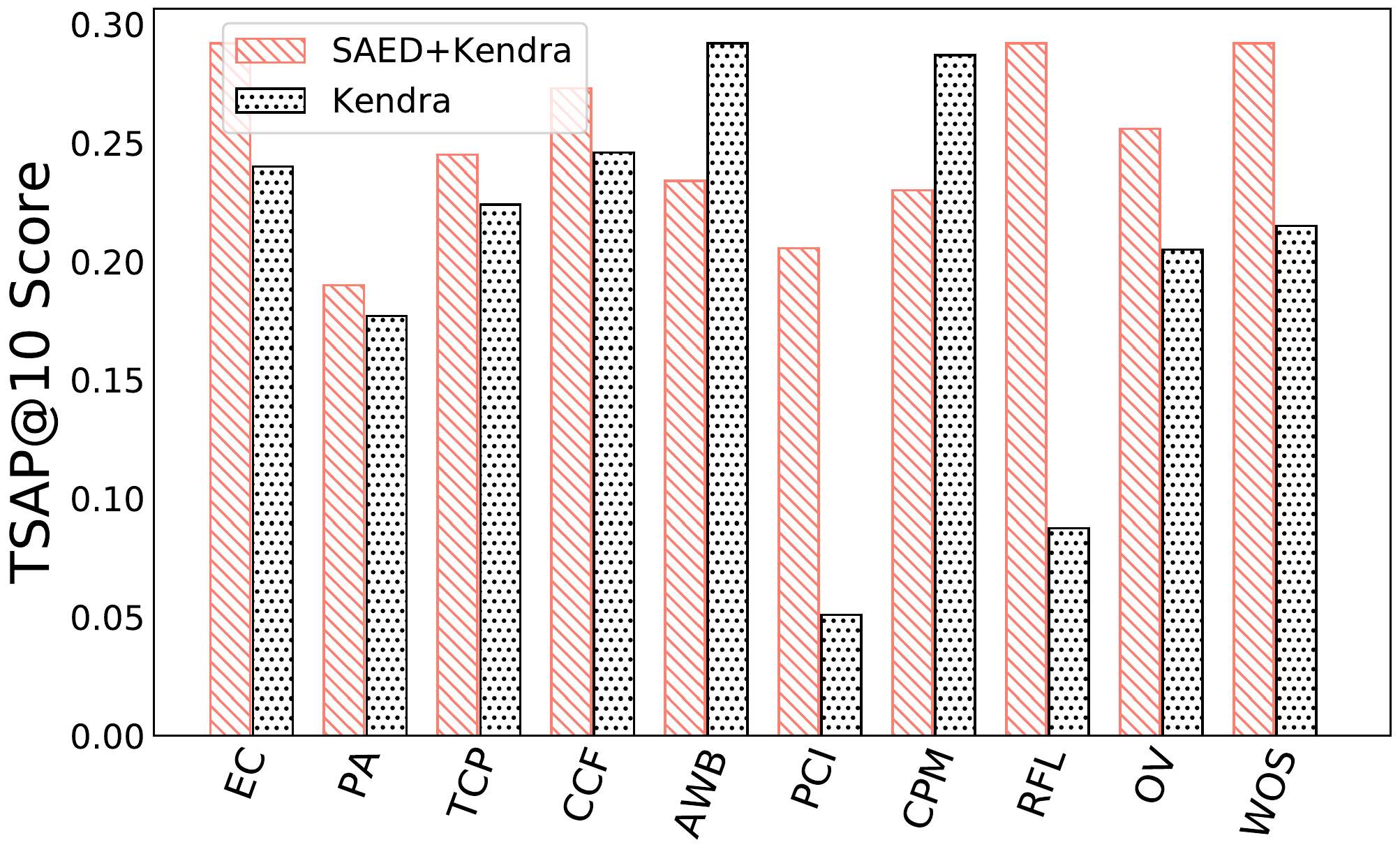}
\caption{BBC dataset}
\label{fig: kendra bbc}
\end{subfigure} \hfill
\begin{subfigure}  {.45\textwidth}
\includegraphics[width=.95\linewidth]{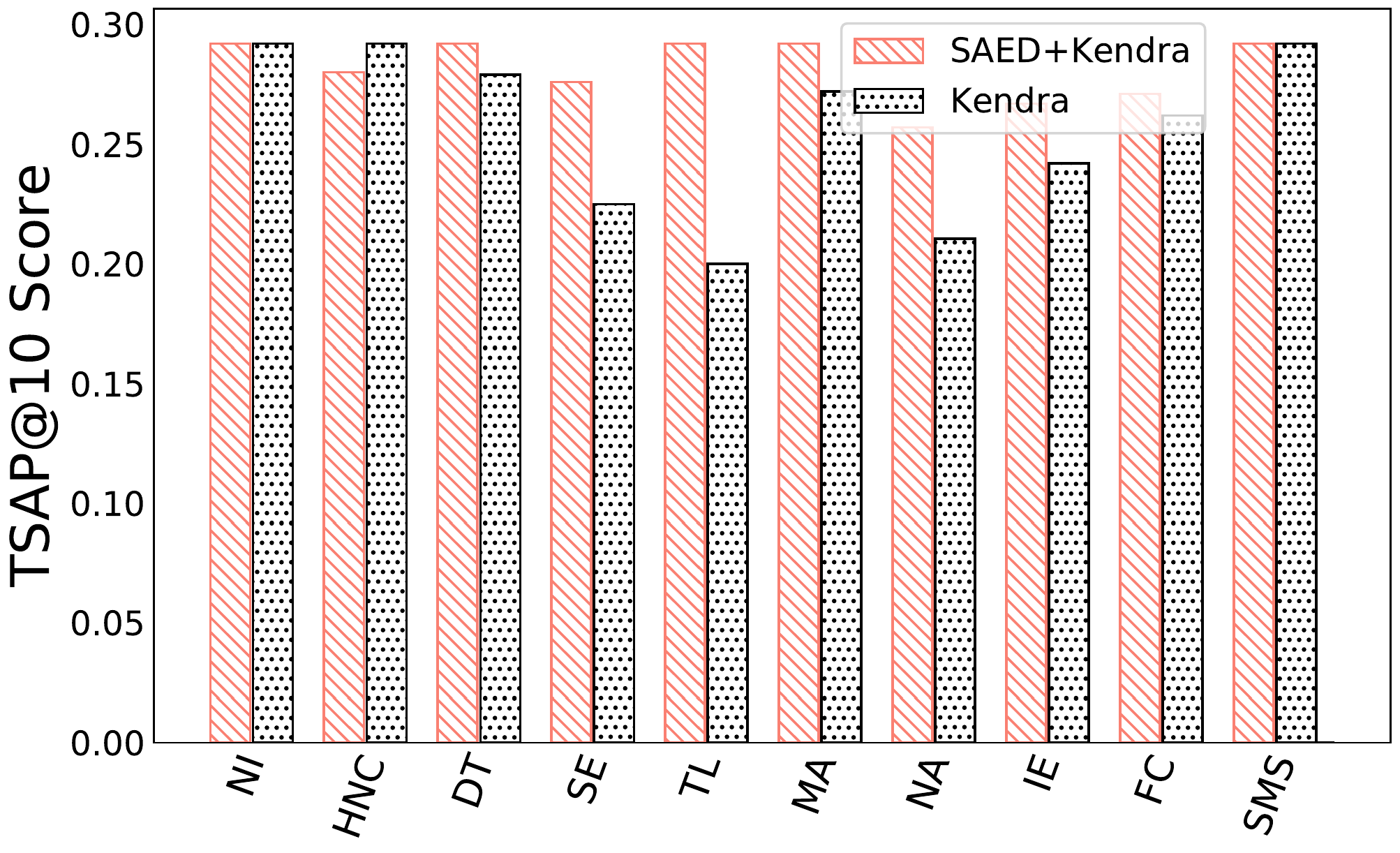}
\caption{RFC dataset} 
\label{fig: kendra rfc}
\end{subfigure} \hfill
\caption{Comparing TSAP@10 scores obtained from \name+Kendra versus AWS Kendra in searching benchmark queries.}
\end{figure*}

In both Figure~\ref{fig: s3bd bbc} and~\ref{fig: s3bd rfc}, we observe that for all queries in both datasets, \name+S3BD outperforms the S3BD system. In addition, we observe that S3BD produces less relevant results for the BBC dataset compared to the RFC dataset. This is because, unlike the RFC dataset, in several cases, the exact keywords of the benchmark queries do not exist in the BBC dataset. The worst case of these issues has occurred for the \texttt{PCI} query in S3BD, because its query expansion procedure could not capture the complete semantics. In contrast, \name+S3BD is able to handle the cases where the exact keyword does not exist in the dataset, thus, we see that it yields to a remarkably higher relevancy.

Even if we consider \texttt{PCI} as an outlier and exclude that from the analysis, in Figure~\ref{fig: s3bd bbc}, we still notice that the TSAP@10 score of \name+S3BD is on average $41.2\%$ higher than S3BD. Although the difference between S3BD and \name+S3BD is less significant for the RFC dataset (in Figure~\ref{fig: s3bd rfc}), we still notice some $17\%$ improvement in TSAP@10 score. This is because RFC is a domain-specific dataset and the exact keywords of queries can be found in the dataset, hence, making use of smart methods to extract the semantic is not acute to earn relevant results. From these results, we can conclude that \name~can be specifically effective for generic datasets where numerous topics exist in the documents.

\paragraph{\textbf{AWS Kendra vs \name+Kendra}}

\begin{figure*}
\begin{subfigure} {.45\textwidth}

\includegraphics[width=.95\linewidth]{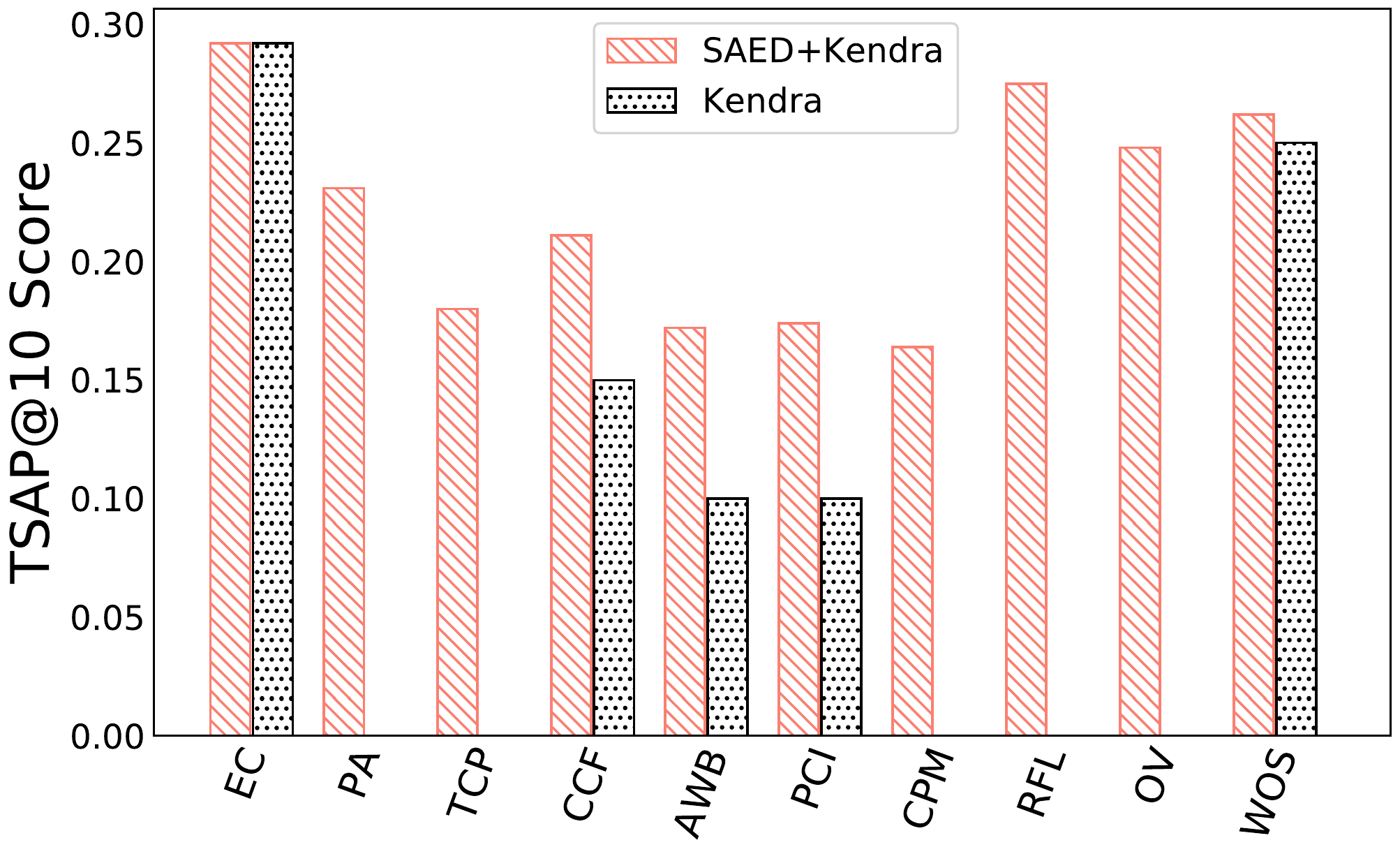}
\caption{Encrypted BBC dataset}
\label{fig: encry kendra bbc}
\end{subfigure} \hfill
\begin{subfigure} {.45\textwidth}

\includegraphics[width=.95\linewidth]{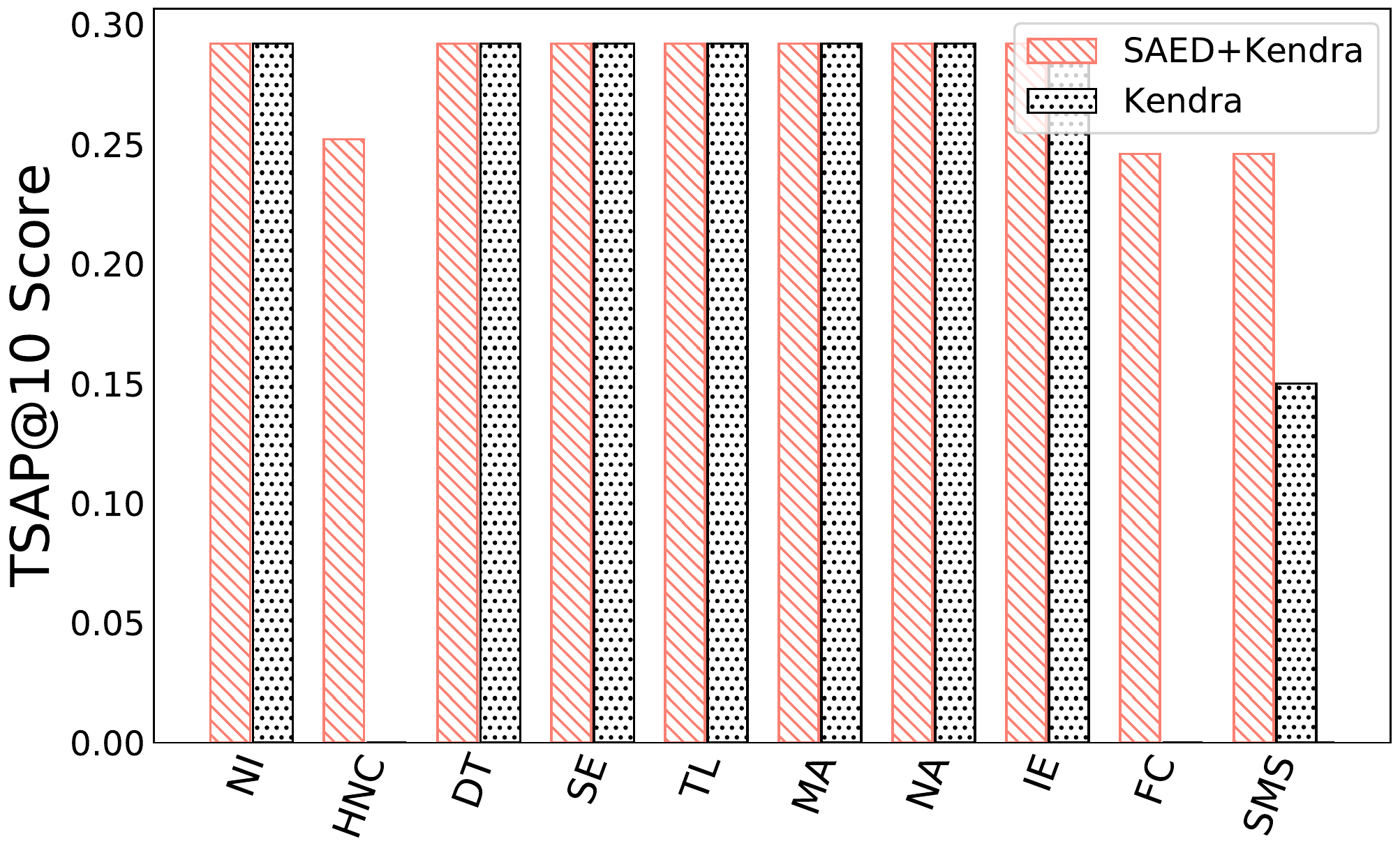}
\caption{Encrypted RFC dataset}
\label{fig: encry kendra rfc}
\end{subfigure} \hfill
\caption{Comparing TSAP@10 scores obtained from \name+Kendra vs AWS Kendra systems in the encrypted domain.}
\end{figure*}

In Figures~\ref{fig: kendra bbc} and \ref{fig: kendra rfc}, we report TSAP@10 score obtained from AWS Kendra versus \name+Kendra for BBC and RFC datasets, respectively.  
Specifically, in Figure~\ref{fig: kendra bbc} (BBC dataset), a significant improvement (on average $26.5\%$) is noticed in the TSAP@10 score of \name+Kendra. However, unlike \name+S3BD, \name+Kendra does not beat Kendra for all the queries. The reason Kendra outperforms \name+Kendra for \texttt{AWB} and \texttt{CPM} queries is that \name~injects extra keywords and sends the expanded query set to AWS Kendra. Then, Kendra returns documents that are related to the queries and to the expanded keywords. We realized that the Ranking unit of \name~occasionally prioritizes documents that include keywords of the expanded queries instead of those with the query keywords. 

Similar to the S3BD experiment, we observe that the relevancy resulted from Kendra and \name+Kendra is less significant for RFC. However, we still obtain around $12\%$ improvement in TSAP@10 score according to Figure~\ref{fig: kendra rfc}.

\subsection{Relevancy of Privacy-Preserving Enterprise Search}

To examine the efficiency of \name~for privacy-preserving enterprise search systems, we conducted experiments using encrypted BBC and RFC datasets. The encrypted datasets were uploaded to the cloud tier and the expanded queries were also encrypted and searched on the cloud tier via Kendra.          

We use the TSAP@10 score, as shown in Figure~\ref{fig: encry kendra bbc} and~\ref{fig: encry kendra rfc}, for the BBC and RFC datasets, respectively. Figure~\ref{fig: encry kendra bbc} indicates that \name+Kendra substantially outperforms Kendra for all the benchmark queries. We can see that for encrypted dataset Kendra cannot do anything except pattern matching and returning documents that exactly include the encrypted query. Therefore, searching for several queries (\eg \texttt{PA},\texttt{TCP}, \texttt {CPM}, etc.) does not retrieve any documents. We notice that, in both systems, the highest TSAP@10 score is in searching \texttt{EC}. The reason is the high number of documents in BBC that contain the exact phrase \texttt{European commission}.

The reported TSAP@10 scores for the RFC dataset in Figure~\ref{fig: encry kendra rfc} shows a clear improvement in compared with the BBC dataset. We observe that seven out of ten queries provide an equal TSAP@10 scores in both systems. The reason that makes Kendra competitive to \name+Kendra is the exact availability of the benchmark queries in RFC. However, for \texttt{HNC} and \texttt{FC}, the exact query keywords are not present in the dataset, hence, Kendra fails to find any results.

\subsection{Discussion of the Relevancy Results}
In Table~\ref{tab:mean score}, we report \emph{mean F-1} and \emph{mean TSAP@10} scores for the \name-plugged enterprise search systems along with their original versions upon utilizing the datasets both in the plain-text and encrypted forms. From the table, we notice that, regardless of the enterprise search system being employed, a higher search relevancy is consistently achieved for the RFC dataset as opposed to the BBC dataset. 

The search relevancy is consistently improved when \name+Kendra is used and it provides on average of $23\%$ improvement in mean F-1 score and $21\%$ in the mean TSAP@10 score. Although original S3BD is the underperformer, using \name+S3BD improves its mean F-1 and mean TSAP@10 scores by
on average of $40\%$ and $32\%$, respectively.

\begin{table}[h]

\centering
\resizebox{\linewidth}{!}{

\begin{tabular}{c|c|c|c|c|}
\cline{2-5}
\textbf{}                                  & \multicolumn{2}{c|}{\textbf{BBC}} & \multicolumn{2}{c|}{\textbf{RFC}} \\ \hline
\multicolumn{1}{|c|}{\textbf{Systems}} &
  \textbf{\begin{tabular}[c]{@{}c@{}}Mean\\ F-1\end{tabular}} &
  \textbf{\begin{tabular}[c]{@{}c@{}}Mean\\ TSAP@10\end{tabular}} &
  \textbf{\begin{tabular}[c]{@{}c@{}}Mean\\ F-1\end{tabular}} &
  \textbf{\begin{tabular}[c]{@{}c@{}}Mean\\ TSAP@10\end{tabular}} \\ \hline \hline
\multicolumn{1}{|c|}{S3BD}                 & 0.50            & 0.17            & 0.80            & 0.24            \\ \hline
\multicolumn{1}{|c|}{SAED+S3BD}            & 0.82            & 0.25            & 0.92            & \textbf{0.28}   \\ \hline
\multicolumn{1}{|c|}{Kendra}               & 0.67            & 0.20            & 0.88            & 0.26            \\ \hline
\multicolumn{1}{|c|}{SAED+Kendra}          & \textbf{0.90}   & \textbf{0.27}   & \textbf{0.93}   & \textbf{0.28}   \\ \hline  
\multicolumn{1}{|c|}{Kendra (Encry.)}      & 0.31            & 0.09            & 0.75            & 0.22            \\ \hline
\multicolumn{1}{|c|}{SAED+Kendra (Encry.)} & 0.73            & 0.22            & 0.90            & 0.27            \\ \hline
\end{tabular}
}
\caption{\small{Comparing the mean F-1 and the mean TSAP@10 scores obtained from \name-plugged enterprise search systems versus their original forms. The highest resulted scores are shown in bold font.}}
\label{tab:mean score}
\end{table}

In the encrypted domain, we notice that \name+Kendra offers a substantially higher (up to $130\%$) search relevancy for BBC dataset. As the exact keywords of the given search queries are not present in the encrypted form of BBC dataset, AWS Kendra fails to perform semantic search, rather does only a pattern matching, which makes it an underperformer for this dataset. 
On the other hand, search relevancy is improved for RFC dataset since mean F-1 and mean TSAP@10 scores are improved by
at least $20\%$. This is because, most of the queries are present exactly in the dataset and Kendra retrieves most of the relevant documents by relying only on pattern matching.

\vspace{-3 pt}
\subsection{Evaluating the Search Time }
\begin{figure} [h]
\centering
\vspace{-6pt}
\includegraphics[width=.90\linewidth]{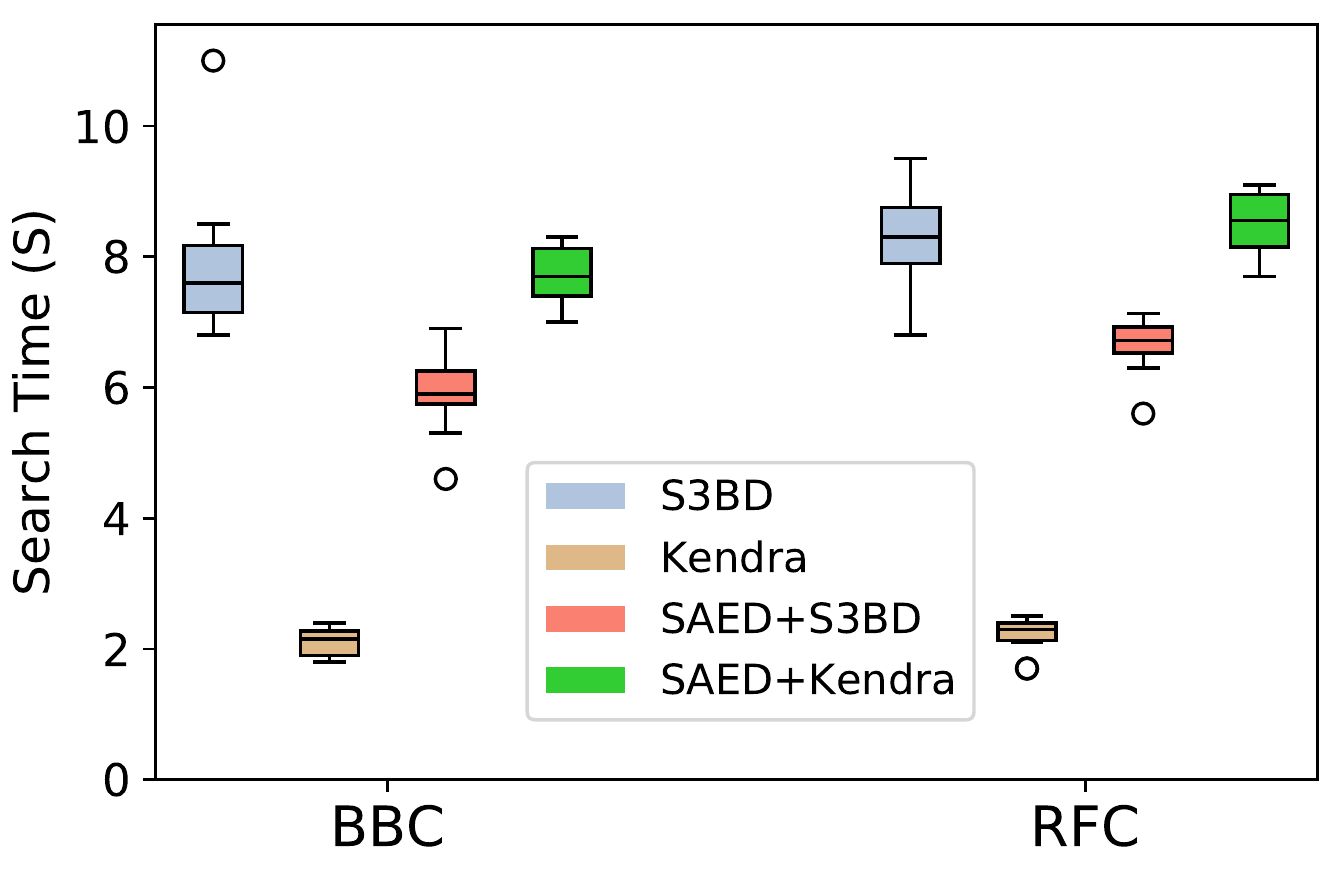}
\caption{Search time comparison among S3BD, Kendra, SAED+S3BD, and SAED+Kendra systems.}
\label{fig: searchtime}
\end{figure}

Figure~\ref{fig: searchtime} presents the total incurred search time of the experimented queries for each dataset. The search time is calculated as the summation of the elapsed time taken by a query to be processed (\eg expansion, weighting) and turnaround time until the result set is received. To eliminate the impact of any randomness in the computing system, we
searched each set of experimented queries 10 times and reported the results in the form of box plots.
The figure indicates that S3BD system has the highest search time overhead for both datasets which could impact real-time searchability in case of big data. \name+S3BD incurs less query processing time overhead compared to the original (unmodified) S3BD system.

On the other hand, AWS Kendra causes the lowest time overhead for both datasets compared to \name+Kendra. \name+Kendra causes around $4$ times more time overhead compared to original Kendra. However, in the prior set of experiments, we determine that \name+Kendra achieves a substantially higher search relevancy for most of the queries and, particularly, for datasets with privacy constraints. 

\section{CONCLUSIONS and Future Work}\label{sec:conclsn}
A context-aware, personalized, and privacy-preserving  enterprise search service is the need of the hour for data owners who wish to use cloud services. Our approach to address this demand was to separate the search intelligence and privacy aspects from the pattern matching aspect. We developed \name~that achieves privacy and intelligence at the edge tier and leaves the large-scale pattern matching for the cloud tier. \name~is pluggable and can work with any enterprise search solution (\eg AWS Kendra and S3BD) without dictating any change on them. Utilizing edge computing on the user's premises preserves the user's privacy and makes \name~a lightweight solution. Leveraging recurrent neural network-based prediction models, WordNet database, and Word2Vec, \name~proactively expands a search query in a proper contextual direction and weights the expanded query set based on the user's interest. In addition, \name~provides the ability to perform semantic search while the data are stored in the encrypted form on the cloud. In this case, the existing enterprise search solutions just perform the pattern matching without knowing the underlying data. Evaluation results, verified by human users, show that \name~can improve the relevancy of the retrieved results by on average $\approx24\%$ for plain-text and $\approx75\%$ for encrypted generic datasets. There are several avenues to improve \name. One avenue is to cover domain-specific and trendy keywords. Another avenue is to make the \name~flexibly deployed on various devices. For instance, when the user is on the move and does not have access to the edge, \name~should shrink to the bare minimum search intelligence and vice versa.

\section*{Acknowledgments}
The research was supported by Chameleon cloud, Amazon
Cloud research credit, and NSF grant IRES-1940619. We are thankful to the students helped creating the benchmark queries and verified the relevancy of the search results.

\bibliographystyle{elsarticle-num}
\linespread{.7}
\bibliography{reference}


\end{document}